\documentstyle[prl,aps,twocolumn,epsf]{revtex}
\def\break#1{\pagebreak \vspace*{#1}}
\begin{document}

\draft

\title{QUANTUM MECHANICAL SPECTRAL ENGINEERING BY 
SCALING INTERTWINING}

\author{DAVID J. FERN\'ANDEZ C.$^{\dagger}$ and HARET C. ROSU$^{\ddagger}$
}

\address{
$^{\dagger}$
{\it Departamento de F\'{\i}sica, CINVESTAV-IPN,
Apdo Postal 14-740, 07000 M\'exico Distrito Federal, Mexico}\\
$^{\ddagger}$
{\it Instituto de F\'{\i}sica de la Universidad de Guanajuato, Apdo Postal
E-143, Le\'on, Guanajuato, Mexico}
}

\maketitle
\widetext

\begin{abstract}
Using the concept of {\it spectral engineering} we explore the
possibilities of building potentials with prescribed spectra offered
by a modified intertwining technique involving operators which are 
the product
of a standard first-order intertwiner and a unitary scaling. In the same 
context
we study the iterations of such transformations finding
that the scaling intertwining provides a different and richer mechanism in 
designing quantum spectra with respect to that given by the standard 
intertwining. 
\end{abstract}

\pacs{PACS number(s): 03.65.Fd, 03.65.Ge, 11.30.Pb   \hfill
Physica Scripta 64 (2001) 177-183}  

\narrowtext
\section{Introduction} The remarkable success in producing artificial 
systems
with controllable physical properties (quantum wells and dots in materials
science \cite{dots}, traps in atomic physics 
\cite{nob98} -\cite{mf89}, spatially confined quantum systems
\cite{ijqc} among others) has stimulated the 
research of the so called quantum engineering (see e.g., the recent review
\cite{har99}). For instance, the control and manipulation of 
effects such as diffraction, widening, interference, etc., of either
specific or arbitrary quantum states have been longly addressed from both
theoretical and experimental viewpoints \cite{fe92} -\cite{brumer}.
In the same spirit, the ``technological" idea of designing potentials with
prescribed quantum spectra \cite{zch97}, a subject that one may 
call {\it spectral engineering} is worth of investigating. Some progress 
in this area can 
be done by restricting the construction to potentials whose spectra are
slightly different from a given initial one through the usage of the well-known
algebraic
procedures provided by the factorization method \cite{ih51} -\cite{fe84},
supersymmetric quantum mechanics \cite{wi81} -\cite{ro95} and other related
techniques \cite{ais93} -\cite{jr98}. It is worth noticing that
these developments can be recovered from the equivalent 
{\it intertwining technique}
\cite{d78}, which provides a better understanding of the generating
process \cite{fe97} -\cite{mi99}. However, nowadays it becomes
apparent the need of novel theoretical techniques additional to the standard
intertwining in order to manufacture quantum spectra and get further 
insight for new technological, device-oriented possibilities at 
nanoscopic and mesoscopic scales. 

With this motivation, we present here a modified intertwining as an
additional mechanism to generate potentials with prescribed spectra. The
modification will consist in replacing the standard intertwiner by an
operator which is the product of a standard intertwiner and a scale
transformation. This constitutes a variant of the method used to analize
the so-called self-similar potentials \cite{sha92,spi92} (for a review see
\cite{spi95}).  Mathematically speaking we will see that this simple
change is \break{1.65in} essentially equivalent to the standard
intertwining. However, we will find that the spectra of the potentials
generated by the scaling intertwining can be manipulated (when the scaling
parameters and the position of the new level are changed) in an
essentially different way from the manipulation possibilities provided by
the standard intertwining. 

The paper is organized as follows. The standard first order intertwining
will be introduced in section 2, while the higher order case as an
iterative procedure of the first order one will be presented in
section 3. In both sections the possibilities for the spectral engineering
offered by each technique will be discussed. The scaling intertwining as a
generalization of the first order method of section 2 will be the subject of
section 4 and its iterations will be addressed in section 5. 
At section 6 the spectral 
possibilities offered by all these techniques will be illustrated by means
of the harmonic oscillator.
Finally, section 7 contains some concluding remarks.


\section{First order intertwining technique} In the standard intertwining
one looks 
for an operator $A_1^{+}$ connecting two Hamiltonians as 
\begin{equation} 
H_1A_1^{+}= A_1^{+}H_0,  
\end{equation} 
where 
\begin{equation}
H_i = -\frac12 \frac{d^2}{dx^2} + V_i(x), \quad i=0,1.
\end{equation}
The first order intertwining uses for $A_1^{+}$ the operator
\begin{equation} 
A_1^{+}=\frac{1}{\sqrt{2}}\left(-\frac{d}{dx}+\alpha_1(x,\epsilon)\right), 
\end{equation} 
leading to the standard Riccati equation for $\alpha_1(x,\epsilon)$
associated to the given initial potential 
\begin{equation} 
\alpha_1'(x,\epsilon) + \alpha_1^{2}(x,\epsilon)=2(V_0(x) -
\epsilon), 
\end{equation} 
where the prime denotes the derivative with respect to $x$. The new potential
$V_1(x)$ is determined according to: 
\begin{equation} 
V_1(x) = V_0(x) - \alpha_1'(x,\epsilon), 
\end{equation}
whenever one is able to find a solution to (4) for given $V_0(x)$ and
$\epsilon$. The {\it factorization energy} $\epsilon\in {\bf R}$ is quite
important for the properties of $V_1(x)$, as it is clear after
substituting $\alpha_1(x,\epsilon) = [\ln\psi(x)]'$ in (4) leading to
\begin{equation}
H_0\psi(x) = -\frac{1}{2}\psi''(x) + V_0(x)\psi(x) = 
\epsilon\psi(x)~.  
\end{equation} 
Although similar to the standard eigenvalue equation for $H_0$, notice
that $\psi(x)$ in (6) is not necessarily normalizable, but it should not
have zeros in order to avoid supplementary singularities of $V_1(x)$ with
respect to those of $V_0(x)$.  As is well known, for $\epsilon > E_0$
($E_0$ is the ground state energy of $H_0$) $\psi(x)$ will always have
zeros. However, if $\epsilon \leq E_0$ it is possible to make $\psi(x)$ to
have no zeros by adjusting the ratio of the two constants in the general
solution of (6), resulting in a physically meaningful $V_1(x)$ as
explained for example by Sukumar \cite{wi81}. The spectrum of
$V_1(x)$ consists of the sequence of levels $E_n$ of the initial
Hamiltonian plus a new `ground state' at $\epsilon$. The scheme becomes
complete after realizing that $H_0$ and $H_1$ are factorized in terms of
$A_1$ and $A_1^+$ as follows 
\begin{equation} 
H_0 = A_1 A_1^+ + \epsilon, \qquad H_1 =
A_1^+A_1 + \epsilon~.  
\end{equation} 
Let us notice that equation (5) is equivalent to $H_1 = H_0
-\alpha_1'(x,\epsilon)$. Thus, the intertwining relationship (1) can be
put in the commutator form $[H_0,A_1^+] = \alpha_1'(x,\epsilon) A_1^+$. 

The possibilities for designing potentials with prescribed spectra offered
by the standard first order intertwining are restricted by our skills to
solve the Riccati equation (4) or its equivalent (6) for different values
of the factorization energy $\epsilon$. A simple situation arises if we
are able to solve (4) for $\epsilon$ sweeping an interval
$I=(\epsilon_{\rm L}, \epsilon_{\rm R})$. By varying $\epsilon$ in such a
domain we are `manipulating' the spectra of the corresponding potentials
$V_1(x)$ in such a way that the excited levels $E_n$ remain fixed but the
ground state level $\epsilon$ is moving with respect to those levels. This
is a modest but interesting contribution to the spectral engineering
offered by the first order intertwining.

\section{Higher order intertwining technique} In order to improve the
spectral controllability, it would be important to explore additional
techniques to find solvable potentials with known spectra. A useful one
consists of iterations of the first order intertwining technique of
section 2 \cite{ais93}-\cite{mi99}. Suppose that, departing from
a solvable Hamiltonian $H_0$, we have constructed a chain of Hamiltonians
of form (2), $\{H_i, i=1,\dots,k\}$, that are pair-intertwined according to
\begin{equation}
H_i A_i^+ = A_i^+ H_{i-1}, \quad i=1,\dots,k,
\end{equation}
where
\begin{equation}
A_i^+ = \frac{1}{\sqrt{2}}\left( -\frac{d}{dx} + \alpha_i(x,\epsilon_i)
\right).
\end{equation}
Thus, $\alpha_i(x,\epsilon_i)$ must satisfy a Riccati equation similar to
(4),
\begin{equation}
\alpha_i'(x,\epsilon_i) + \alpha_i^2(x,\epsilon_i) = 2[V_{i-1}(x) -
\epsilon_i]~,
\end{equation}
and $V_i(x)$ is related to $V_{i-1}(x)$ according to
\begin{equation}
V_i(x)  = V_{i-1}(x) - \alpha_i'(x,\epsilon_i)~.
\end{equation}
It is simple to show that $\alpha_i(x,\epsilon_i)$ can be obtained from
$\alpha_{i-1}$ evaluated at two different factorization energies
$\epsilon_i, \ \epsilon_{i-1}$ through the following finite difference
formula \cite{fe97}-\cite{mi99}: 
\begin{equation}
\alpha_i(x,\epsilon_i) = -\alpha_{i-1}(x,\epsilon_{i-1}) - 
\frac{2(\epsilon_{i-1} - \epsilon_i)}{\alpha_{i-1}(x,\epsilon_{i-1}) -
\alpha_{i-1}(x,\epsilon_{i})}~.
\end{equation}
Let us notice that this equation was previously obtained by Adler when
studying B\"acklund transformations for Painlev\'e equations \cite{ad93}. 
By iterating this formula in the subindex $i$ it is clear that the final
potential $V_k(x) = V_{k-1}(x) - \alpha_k'(x,\epsilon_k)$ is determined by
$k$ solutions $\alpha_1(x,\epsilon_i), \ i=1, \dots,k$ to the initial
Riccati equation (4) associated to the $k$ factorization energies
$\epsilon_1, \dots, \epsilon_k$. The spectrum of $V_k(x)$ consists of the
levels $E_n$ of the original Hamiltonian $H_0$ plus $k$ new levels below
the ground state energy level $E_0$ of $H_0$ (by simplicity ordered
according to $\epsilon_{k} < \epsilon_{k-1}<\cdots<\epsilon_1< E_0$). 

As in the first order intertwining, in the higher order case the spectral
engineering depends on how many solutions we are able to find for the
initial Riccati equation (4) as well as on the positions of the $k$
different factorization energies $\epsilon_1,\dots,\epsilon_k$. If we are
able to find solutions only for isolated values of
$\epsilon_1,\dots,\epsilon_k$, we can hardly manipulate the spectrum of
the potentials $V_k(x)$. The really interesting situation arises if we are
able to find solutions for $\epsilon$ sweeping an interval $I$. Let us
choose $k$ of these solutions $\alpha_1(x,\epsilon_i), \ i=1,\dots,k$.  By
moving those $k$ values in such a way that always $\epsilon_i\in I$ and
the previous ordering of the $\epsilon_i$'s is respected, we will be
manipulating the spectrum of $V_k(x)$ so that the levels $E_n$ are fixed
but the $k$ ones $\epsilon_i, \ i =1,\dots, k$ will be moving in a way
that could be selected according to some specific purposes. This provides
more freedom for controlling quantum spectra than that offered by the first 
order intertwining. Moreover, we shall see below that it is possible to design
different tools based on the same mathematical formalism of section 2
which will provide new mechanisms to manipulate quantum spectra. 

\section{The scaling intertwining method} Let us consider a slight
generalization of the first-order intertwining of section 2 by maintaining
Eq.~(1) but introducing now a `deforming' parameter $q_1$ into $H_1$ as
$H_1= q_1^2H_0 - f(x)$, and substituting the intertwining operator (3) by
the same standard intertwiner times a scaling unitary operator
$U_1 \equiv e^{i\frac{\lambda_1}{2}(\hat x\hat p+\hat p\hat x)} =
e^{\frac{\lambda_1}{2}(x\frac{d}{dx}+\frac{d}{dx}x)}, \ \lambda_1\in{\bf
R}$ (see, e.g., \cite{sha92}-\cite{nnr99}):
\begin{equation}  
A_{\lambda_1}^{+}=\frac{1}{\sqrt{2}}\left(-\frac{d}{dx} +
\alpha_1(x,\epsilon_1)\right)U_1.
\end{equation}
Multiplying Eq.~(1) by $U_1^{+}$ to the right one gets
$$ 
\big[q_1^2H_0-f(x)\big]\left[-\frac{d}{dx}+\alpha_1(x,\epsilon_1) \right]
$$
\begin{equation}
=\left[-\frac{d}{dx} +\alpha_1(x,\epsilon_1)\right]U_1H_0 U_1^{+},
\end{equation}
and making equal on both sides of (14) the coefficients of the various
corresponding powers of the derivative operator $\frac{d}{dx}$ leads to
\begin{equation} 
 q_1^2 = e^{-2\lambda_1},
\end{equation}
\begin{equation} 
q_1^2\alpha_1'(x,\epsilon_1)+ q_1^2V_0(x)-f(x)=V_0(e^{\lambda_1} x), 
\end{equation}
$$
 -\frac{q_1^2}{2}\alpha_1''(x,\epsilon_1)+ [q_1^2V_0(x) -
f(x)]\alpha_1(x,\epsilon_1)
$$
\begin{equation}
=\alpha_1(x,\epsilon_1)V_0(e^{\lambda_1} x)-V_0'(e^{\lambda_1} x) .
\end{equation}
After substituting $\lambda_1 = -\ln q_1$, Eq.~(16) reads
\begin{equation} 
f(x)=q_1^2\alpha_1'(x,\epsilon_1) + q_1^2V_0(x)-V_0(x/q_1)~.
\end{equation}
Notice the non-local character of equation (18) for $q_1\neq 1$ in
contrast with the local nature of the standard intertwining ($q_1=1$) for
which $f(x) = \alpha_1'(x,\epsilon_1)$:  in the scaling intertwining the
original potential at two different points, $x$ and $x/q_1$, interfere
with $\alpha_1'(x,\epsilon_1)$ in order to provide $f(x)$.  This
difference arises even when $V_0(x)$ is a homogeneous function of degree
$d$ for which (18) reduces to the more local form
$f(x)=q_1^2\alpha_1'(x,\epsilon_1) + (q_1^2-q_1^{-d})V_0(x)$. When $d=-2$
one gets $f(x)=q_1^2\alpha_1'(x,\epsilon_1)$ which is the closest one can
reach to the standard intertwining.  Substituting $\lambda_1 = -\ln q_1$
and (18) in Eq.~(17) and integrating it we get the Riccati equation
\begin{equation} 
q_1^2\alpha_1'(x,\epsilon_1) + q_1^2\alpha_1^{2}(x,\epsilon_1) =
2[V_0(x/q_1)-\epsilon_1]~,
\end{equation}
where, as it will be clear below, it is convenient to choose the
integration constant equal to one of the factorization energies
$\epsilon_1$ of (4).  Now, if one makes the change of variable $y=x/q_1$
and denotes $\tilde{\alpha}_1(y,\epsilon_1) \equiv
q_1\alpha_1(q_1y,\epsilon_1)$ one easily gets
\begin{equation} 
\frac{d\tilde{\alpha}_1(y,\epsilon_1)}{dy}+ 
\tilde{\alpha}_1^{2}(y,\epsilon_1)=
2(V_0(y) - \epsilon_1).
\end{equation}
This is again the Riccati equation (4) with factorization energy
$\epsilon_1$.  Therefore, the same Riccati solution used in the standard
intertwining (Eqs.~(1) to (7)) can be used in the scaling
intertwining of Eqs.(1,13-20) as well in order to generate solvable potentials
with known spectra. The eigenfunctions of $\tilde H_1\equiv q_1^{-2} H_1=
H_0 - q_1^{-2} f(x)$ are proportional to the action of $A_{\lambda_1}^+$
on the eigenfunctions of $H_0$ due to the fact that the two Schr\"odinger
Hamiltonians $H_0$ and $\tilde H_1$ are intertwined in the slightly
generalized way 
\begin{equation}
\tilde H_1 A_{\lambda_1}^{+}= q_1^{-2} A_{\lambda_1}^{+} H_0~.
\end{equation}
The potential associated to $\tilde H_1$ takes the form
\begin{equation}
\tilde V_1(x) = V_0(x) - q_1^{-2} f(x) = q_1^{-2} V_0(x/q_1) -
\alpha_1'(x,\epsilon_1)~. 
\end{equation}
The corresponding eigenvalues are $\{q_1^{-2}\epsilon_1, q_1^{-2}E_n\}$,
where $q_1^{-2}\epsilon_1$ is the ground state energy of $\tilde H_1$
associated to the eigenfunction $\tilde\psi_{\epsilon_1}^{(1)}(x)  \propto
\exp(-\int_0^x\alpha_1(\xi,\epsilon_1) d\xi)$. The factorizations of $H_0$
and $\tilde H_1$ in terms of $A_{\lambda_1}$ and $A_{\lambda_1}^+$ become
\begin{equation}  
H_0 = q_1^2A_{\lambda_1} A_{\lambda_1}^+ + \epsilon_1, \qquad \tilde H_1
= A_{\lambda_1}^+A_{\lambda_1} + q_1^{-2}\epsilon_1~.
\end{equation}

The spectral engineering skills depend once again on the kind of solutions
to the Riccati equations (4) and (20) we are able to get. If we find a solution
for an isolated value of the factorization energy $\epsilon_1$, the
potentials $\tilde V_1(x)$ will have the spectrum $\{q_1^{-2}\epsilon_1,
q_1^{-2} E_n\}$, and by varying $q_1$ we will be just scaling the basic
spectrum $\{\epsilon_1, E_n\}$ at $q_1=1$ generated by means of the
standard intertwining when a new level is created at $\epsilon_1$ starting
from $V_0(x)$. However, notice that solutions of (20)  for $\epsilon_1$
sweeping a real interval $I=(\epsilon_{\rm L}, \epsilon_{\rm R})$, where
$\epsilon_{\rm L}$ and $\epsilon_{\rm R}$ have the same sign, suggest 
choosing $\epsilon_1 = q_1^2 \epsilon_{1\rm F}$ with $\epsilon_{1\rm F}\in
I$ fixed, and varying $q_1$ in such a way as to keep $\epsilon_1\in I$. Thus,
we will manipulate the spectrum of $\tilde V_1(x)$, $\{\epsilon_{1\rm
F}, q_1^{-2} E_n\}$ by scaling by a factor $q_1^{-2}$ the excited levels $E_n$ 
while the ground state energy remains
fixed. This effect is opposite to that produced by the standard
intertwining where the excited levels are static while the changing one is
the ground state level. Thus, we have at hand two different tools
increasing our skills in the spectral engineering that could be
useful in various physical situations.  Moreover, as we shall see below
the scaling intertwining can be iterated in a similar way as the standard
intertwining. 

In order to implement the iterations of this technique, let us notice that
the Hamiltonian $H_1 = q_1^2 H_0 - f(x)$ expressed in the coordinate
$y=x/q_1$ acquires the standard Schr\"odinger form
\begin{equation}
H_1 = - \frac12\frac{d^2}{dy^2} + V_1(y),
\end{equation}
where 
\begin{equation}
V_1(y) \equiv q_1^2 \tilde V_1(q_1y) = V_0(y) - \frac{d\tilde\alpha_1
(y,\epsilon_1)}{dy}.
\end{equation}
These formulae clarify better the scaling intertwining mechanism: the
modified intertwiner $A_{\lambda_1}^+$ connects two almost isospectral
Hamiltonians $H_0$ and $H_1$, creating a new level for $H_1$ at
$\epsilon_1$ and simultaneously changing the coordinate $x$ to $y=x/q_1$.
When we come back to the original coordinate $x$ and scale the energy, we 
arrive at the Hamiltonian $\tilde H_1$ having the standard Schr\"odinger form 
in terms of $x$. In this way we get the modified intertwining of Eq.~(21) which 
leads to the spectral properties previously discussed.
 
\section{Iterative scaling intertwining} Let us restrict our considerations
to two iterations of the scaling intertwining. The generalization
to higher iterative orders follows straightforwardly from the pair of
transformations we shall study in this section. 
During the first step we will have $H_1
A_{\lambda_1}^+ = A_{\lambda_1}^+ H_0$, as in the previous section, while
in the second step we will have $H_2 A_{\lambda_2}^+ = A_{\lambda_2}^+
H_1$.  Taking into account the remark at the end of section 4, it is clear
that in the second step the Hamiltonian $H_1$ of Eq.~(24) `is changed'
into a new one $H_2$ expressed in a different coordinate $z=y/q_2$ as
follows
\begin{equation}
H_2 = - \frac12\frac{d^2}{dz^2} + V_2(z)~.
\end{equation}
The corresponding scaling intertwining operator is now:
\begin{equation}  
A_{\lambda_2}^{+}=\frac{1}{\sqrt{2}}\left(-\frac{d}{dy} +
\alpha_2(y,\epsilon_2)\right)U_2~,
\end{equation}
where $U_2 \equiv e^{i\frac{\lambda_2}{2}(\hat x\hat p+\hat p\hat x)} =
e^{\frac{\lambda_2}{2}(y\frac{d}{dy}+\frac{d}{dy}y)}$. The appropriate
reading of equations (14-18) leads to $\lambda_2 = -\ln q_2$ and an
equation analogue to (19) 
\begin{equation} 
q_2^2\frac{d\alpha_2(y,\epsilon_2)}{dy} + q_2^2\alpha_2^{2}(y,\epsilon_2)
= 2[V_1(y/q_2)-\epsilon_2]~.
\end{equation}
With the change of variables $z=y/q_2$ and $\tilde\alpha_2(z,\epsilon_2) 
\equiv q_2\alpha_2(q_2 z,\epsilon_2)$ we arrive at 
\begin{equation} 
\frac{d\tilde{\alpha}_2(z,\epsilon_2)}{dz}+
\tilde{\alpha}_2^{2}(z,\epsilon_2)=
2(V_1(z) - \epsilon_2),
\end{equation}
where $V_1(z)$ is given by (25). A direct use of equations (10-12) with $i=2$ 
leads to a simple expression for the solution to Eq.~(29) in terms of two
solutions to (4) and (20) at the factorization energies $\epsilon_1$ and
$\epsilon_2$, respectively
\begin{equation} 
\tilde{\alpha}_2(z,\epsilon_2) = - \tilde{\alpha}_1(z,\epsilon_1) - 
\frac{2(\epsilon_1-\epsilon_2)}{\tilde{\alpha}_1(z,\epsilon_1) -
\tilde{\alpha}_1(z,\epsilon_2)}.
\end{equation}
According to (25), the potential $V_2(z)$ reads
$$
V_2(z) = V_1(z) - \frac{d\tilde\alpha_2(z,\epsilon_2)}{dz} 
$$
\begin{equation}
=V_0(z) + 
\frac{d}{dz}
\left[\frac{2(\epsilon_1-\epsilon_2)}{\tilde{\alpha}_1(z,\epsilon_1) -
\tilde{\alpha}_1(z,\epsilon_2)}
\right].
\end{equation}
It has the same levels $E_n$ as $V_0(x)$ plus two additional ones at
$\epsilon_1$ and $\epsilon_2$ ordered by simplicity as
$\epsilon_2<\epsilon_1<E_0.$ We notice that Samsonov has recently shown 
the possibility to construct physically relevant potentials (without
singularities) by using a second order intertwining technique with
$\epsilon_1>E_0$ and $\epsilon_2>E_0$ \cite{sam99}. It is a simple 
matter
now to construct a potential such that the corresponding Hamiltonian
$\tilde H_2 = (q_1q_2)^{-2}H_2$ acquires the standard Schr\"odinger form
in the original coordinate $x$ 
\begin{equation}
\tilde H_2 = - \frac12\frac{d^2}{dx^2} + \tilde V_2(x)~,
\end{equation}
$$
\tilde V_2(x) = \frac1{(q_1q_2)^{2}}V_2 \left(\frac{x}{q_1q_2}\right) =
\frac1{(q_1q_2)^{2}}V_0\left(\frac{x}{q_1q_2}\right) 
$$
\begin{equation}
+\frac{1}{q_1q_2}\left[
\frac{2(\epsilon_1-\epsilon_2)}{\tilde{\alpha}_1(\frac{x}{q_1q_2},\epsilon_1)
- \tilde{\alpha}_1(\frac{x}{q_1q_2},\epsilon_2)}
\right]'~.
\end{equation}
The generalized intertwining relationship involving $H_0$ and $\tilde H_2$
is the following
\begin{equation}
\tilde H_2 \left(A_{\lambda_2}^+ A_{\lambda_1}^+\right) = (q_1q_2)^{-2}
\left(A_{\lambda_2}^+A_{\lambda_1}^+\right) H_0.
\end{equation}
Thus, the spectrum of $\tilde H_2$ is of the form
$\{(q_1q_2)^{-2}\epsilon_2, (q_1q_2)^{-2}\epsilon_1,(q_1q_2)^{-2}E_n, n =
0,1,\dots\}$.

As we have been discussing along this paper, if we get solutions to the
Riccati equations (4) and (20) for isolated values of $\epsilon_1$ and
$\epsilon_2$, when varying the product $q_1q_2$ (by changing either $q_1$, or 
$q_2$, or both) in order to manipulate the spectra of $\tilde V_2(x)$ we will be
just scaling the basic spectrum $\{\epsilon_2, \epsilon_1,E_n\}$ obtained
from $H_0$ by performing two iterations of the standard intertwining with
factorization energies $\epsilon_1$ and $\epsilon_2$. Once again, from the
spectral engineering viewpoint the most interesting cases arise when we
find solutions for $\epsilon_1$ and $\epsilon_2$ sweeping a real interval
$I$. If this would be the case, by choosing for instance $\epsilon_1 =
q_1^2 \epsilon_{1\rm F}$, where the three parameters $\epsilon_{1\rm F}\in
I$, $q_2$ and $\epsilon_2$ are fixed, when varying $q_1$ so that always
$\epsilon_1\in I$ and the ordering between $\epsilon_1$ and $\epsilon_2$
is maintained ($\epsilon_2<q_1^2 \epsilon_{1\rm F}$), the unscaled
spectrum $\{(q_2^{-2}\epsilon_2), (q_2^{-2}\epsilon_{1\rm F}),
(q_2^{-2}E_n)\}$ at $q_1=1$ will become scaled by the factor $q_1^{-2}$
except for the first excited level ($q_2^{-2}\epsilon_{1\rm F}$) that will
remain fixed.  A similar static effect onto the ground state level can be
obtained if we choose $\epsilon_2 = q_2^2 \epsilon_{2\rm F}$, where now the
three fixed parameters are $\epsilon_{2\rm F}$, $q_1$ and $\epsilon_1$: by
changing $q_2$ so that $q_2^2 \epsilon_{2\rm F}<\epsilon_1$ the excited
levels at $q_2=1$ are scaled by the factor $q_2^{-2}$ while the ground
state level remains fixed, as can be seen from the spectrum of $\tilde
V_2(x)$, $\{ (q_1^{-2}\epsilon_{2\rm F}), q_2^{-2} (q_1^{-2}\epsilon_1),
q_2^{-2} (q_1^{-2}E_n)\}$. A third interesting possibility arises by
choosing $\epsilon_1 = (q_1q_2)^2 \epsilon_{1\rm F}$ and $\epsilon_2 =
(q_1q_2)^2 \epsilon_{2\rm F}$ with $\epsilon_{1\rm F}$ and $\epsilon_{2\rm
F}$ fixed:  when varying the product $q_1q_2$ by maintaining always
$\epsilon_1\in I$ and $\epsilon_2\in I$, we will be basically scaling the
spectrum $\{\epsilon_{2\rm F}, \epsilon_{1\rm F}, E_n\}$ at $q_1=q_2=1$ by
the factor $(q_1q_2)^{-2}$ except for the two fixed lowest levels
$\epsilon_{2\rm F}$ and $\epsilon_{1\rm F}$, i.e., $Spectrum[\tilde H_2] =
\{\epsilon_{2\rm F}, \epsilon_{1\rm F}, (q_1q_2)^{-2}E_n\}$.

\section{Example: the harmonic oscillator} Let us illustrate the spectral
engineering effects previously discussed through the simple example of the
harmonic oscillator potential $V_0(x) = x^2/2$. The key point is
to find the general solution to the Riccati equation (4) for different
values of the factorization energy $\epsilon$. Let us notice that the case
with $\epsilon = -1/2$ was solved for the first time by Mielnik
\cite{mi84}. For an arbitrary value of $\epsilon<1/2$, the general
solution has been found by Sukumar \cite{wi81}, and a convenient
expression in terms of confluent hypergeometric functions has been
provided by Junker and Roy \cite{jr98} (see also \cite{fe97}) 
$$
\alpha(x,\epsilon) = x + \frac{d}{dx}\bigg\{\!\ln \bigg[
{}_1F_1\bigg(\frac{1+2\epsilon}{4},\frac12;-x^2\bigg) 
$$
\begin{equation}
+2\nu\frac{\Gamma(\frac{3 -
2\epsilon}{4})}{\Gamma(\frac{1-2\epsilon}{4})} \, x \,
{}_1F_1\bigg(\frac{3+2\epsilon}{4},\frac32;-x^2\bigg)
\bigg]\bigg\}.
\end{equation}
We have denoted the solution of (4) as $\alpha(x,\epsilon)$ instead of
$\alpha_1(x,\epsilon)$ because this will simplify the discussion for all
the intertwinings of this paper.  We can distinguish four different
possibilities of the spectral engineering.

\medskip

\noindent {\bf (i).} As it was discussed in section 2, the potentials
\begin{equation}
V_1(x) = \frac{x^2}{2} - \alpha_1'(x,\epsilon_1),
\end{equation}
$\alpha_1(x,\epsilon_1)\equiv\alpha(x,\epsilon_1)$ as
given by (35), have spectra $\{\epsilon_1,E_n = n + 1/2, \ n =
0,1,\cdots\}$. In order to avoid singularities in $V_1(x)$ we must have
$\vert\nu_1\vert <1$, where $\nu_1$ denotes the $\nu$-constant of (35) 
associated to the factorization energy $\epsilon_1$. The simplest spectral
manipulation consists in varying $\epsilon_1$ in the domain
$(-\infty,1/2)$ so that the ground state level of $V_1(x)$ is changing
while the excited levels $E_n$ remain fixed \cite{jr98}.  This kind
of spectral engineering is illustrated in figure 1 for $\nu_1 =0.9$:  if
the ground state energy of $H_1$ is moved, the potentials (36) have to
change their form in order to maintain the excited levels $E_n$ at fixed
positions. It is worth to notice that the potentials we are drawing are
non-symmetric ($V_1(-x)  \neq V_1(x)$). The symmetric case arises for
$\nu_1 = 0$ (see \cite{jr98}).

\hspace*{2truecm}
\begin{center}
\epsfxsize=9truecm
\epsfbox{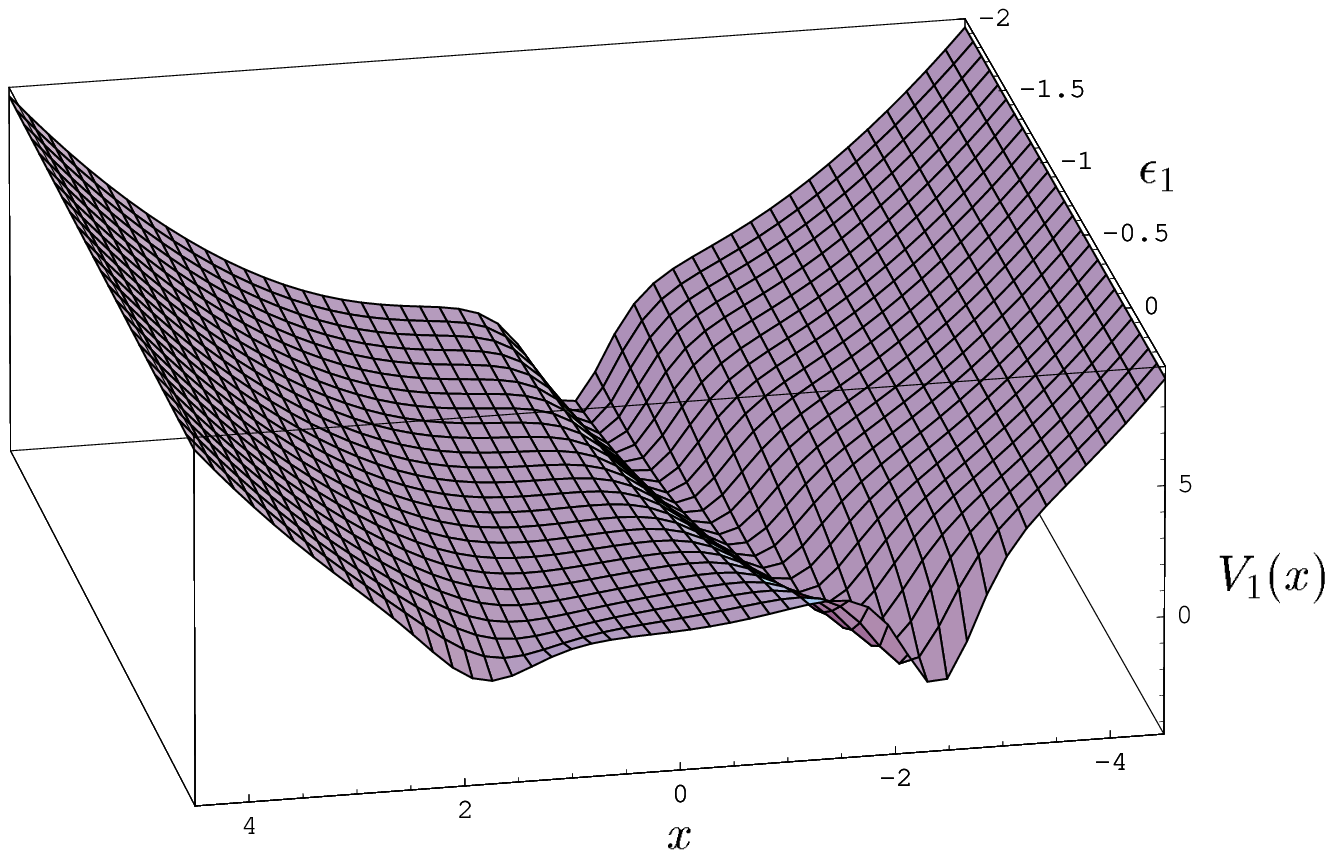}
\end{center}
\medskip
\begin{center}
{\small FIG. 1. 
Spectral manipulation of the potentials $V_1(x)$ of (36) as the ground
state energy $\epsilon_1$ moves inside $[-2,\frac12)$ and $\nu_1=0.9$. The
excited levels $\{E_n=n+ \frac12,n=0,1,\dots\}$ remain fixed for any
$\epsilon_1\in (-\infty,\frac12)$.
} 
\end{center}

\medskip

\noindent {\bf (ii).} Consider now the iterations of the first order
intertwining of section 3. In order to simplify the discussion, we
restrict ourselves to two iterations. The potentials at the end of those
two transformations turn into
\begin{equation}
V_2(x) = \frac{x^2}{2} + \left[
\frac{2(\epsilon_1-\epsilon_2)}{\alpha_1(x,\epsilon_1) -
\alpha_1(x,\epsilon_2)} \right]', 
\end{equation}
where once again $\alpha_1(x,\epsilon_1) \equiv \alpha(x,\epsilon_1)$ and
$\alpha_1(x,\epsilon_2) \equiv \alpha(x,\epsilon_2)$. Notice that to each
factorization energy $\epsilon_i$ an arbitrary
constant $\nu_i, i=1,2$ is associated (see (35)). If the two new levels are 
ordered as
$\epsilon_2<\epsilon_1<1/2$, the domains of $\nu_1$ and $\nu_2$ have to be
restricted to $\vert\nu_1\vert<1$ and $\vert\nu_2\vert>1$ in order to
avoid singularities in $V_2(x)$.

The spectral engineering can be implemented by varying $\epsilon_2$,
$\epsilon_1$, or both. If we sweep $\epsilon_2$ and maintain fixed
$\epsilon_1$, we obtain again the kind of manipulation described at issue
{\bf (i)} consisting in fixing the excited levels of $V_2(x)$ and moving
only the ground state energy. A physically different situation arises if
$\epsilon_2$ is static but $\epsilon_1$ changes, i.e., the first
excited level of $V_2(x)$ is moving. This kind of spectral manipulation is
illustrated in figure 2 for $\epsilon_1 \in[-0.4,0.4], \ \epsilon_2 =
-1/2, \ \nu_1=0$ and $\nu_2 = 10000$. The fixed levels are placed at $\{
-1/2, 1/2,3/2,\dots\}$. Notice the almost symmetric nature of $V_2(x)$
($V_2(-x) \approx V_2(x)$); the strict symmetry is achieved for $\nu_1 =
0$ and $\vert\nu_2\vert\rightarrow \infty$. We also notice the
triple well feature, characteristic of some of the potentials derived by the
iteration of two first order intertwinings. For higher orders of iteration
it is possible to design more complicated multiple well potentials. 
As far as we know, this is the first time that potentials where the first
excited level is movable have been derived. Moreover, a more general
manipulation procedure can be designed by changing $\epsilon_1$ and
$\epsilon_2$ independently to each other.  In such a case, the fixed part
of the spectrum $\{3/2,5/2,\dots\}$ will start from the second excited
state while the first two energy levels will be moving according to the
choice of the spectral manipulator. 

\hspace*{2truecm}
\begin{center}
\epsfxsize=9truecm
\epsfbox{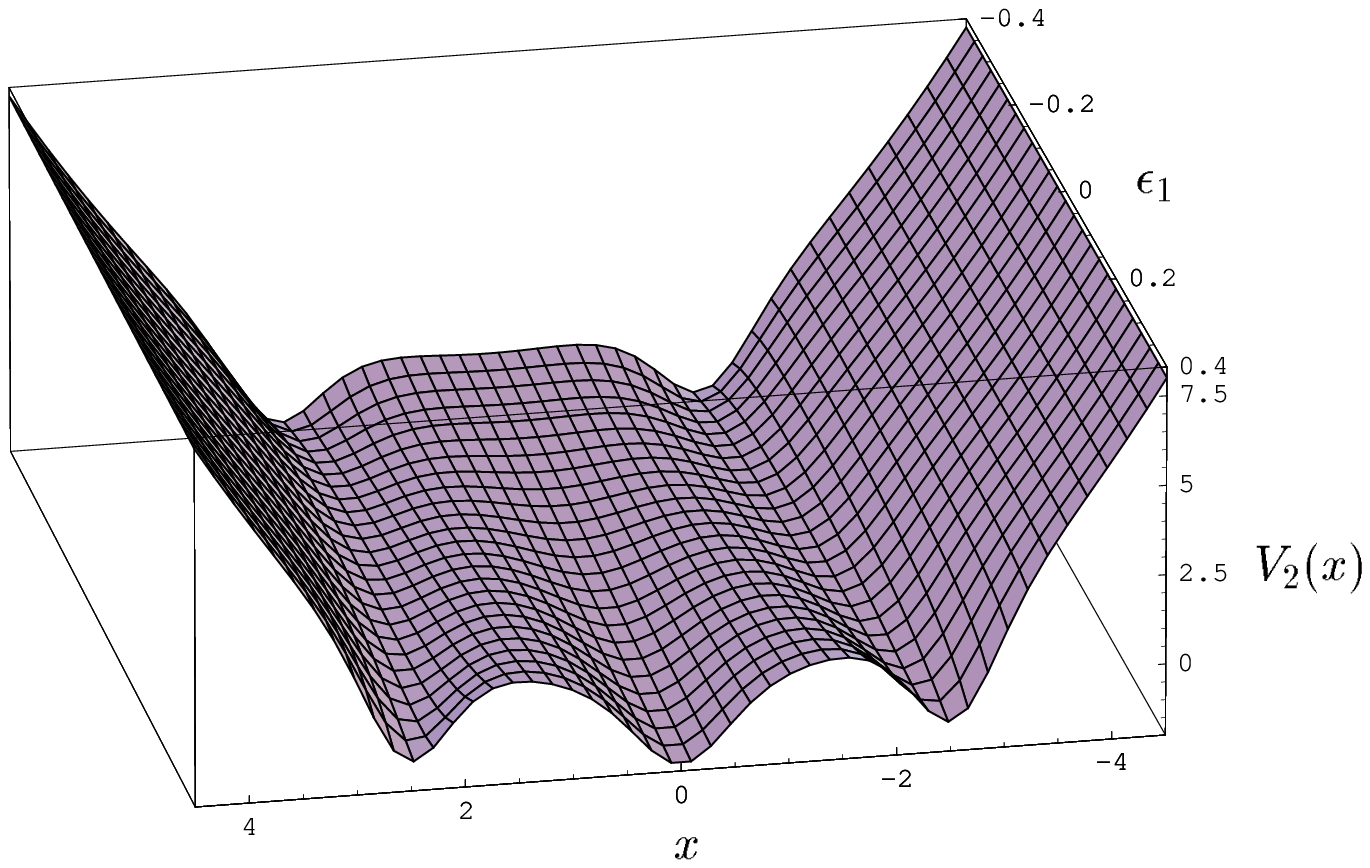}
\end{center}
\medskip
\begin{center}
{\small FIG. 2.
Spectral manipulation of the potentials $V_2(x)$ of (37) as the first
excited level $\epsilon_1$ moves inside $[-0.4,0.4]$ and $\nu_1=0, \
\nu_2 = 10000$. The fixed part of the spectrum consists of the levels
$\{-\frac12,\frac12,\frac32,\dots\}$. 
}
\end{center}

\medskip

\noindent {\bf (iii).} Let us analyze now the spectral engineering
possibilities offered by the scaling intertwining of section 4. The
potentials generated by means of this technique from the oscillator
potential read:
\begin{equation} 
\tilde V_1(x) = q_1^{-4}\frac{x^2}{2} - q_1^{-1} 
\tilde\alpha_1'(x/q_1,\epsilon_1), 
\end{equation}
where $\tilde\alpha_1(x,\epsilon_1) \equiv \alpha(x,\epsilon_1)$. The
spectrum of $\tilde V_1(x)$ is of the form $\{q_1^{-2}\epsilon_1,
q_1^{-2}(n + 1/2), n=0,1,\dots\}$. By taking
$\epsilon_1=q_1^{2}\epsilon_{1\rm F}\in I_1=(-\infty,0)$, 
$\epsilon_{1\rm F}\in I_1$ is fixed, and varying the scaling parameter 
$q_1$, the spectrum
of $\tilde V_1(x)$ will be changing as $\{\epsilon_{1\rm F}, q_1^{-2}(n +
1/2), n=0,1,\dots\}$, i.e., the (excited) levels $E_n=n+1/2$ at $q_1=1$
will be scaled by the factor $q_1^{-2}$, but the ground state level will
remain fixed. A similar treatment can be implemented for $\epsilon_1\in
I_2 = (0,1/2)$. A representation of this kind of spectral manipulation on
$\tilde V_1(x)$ as $q_1$ is changed, with $ \epsilon_{1\rm F}= -1/2$ and
$\nu_1=0$, is shown in figure 3. Notice that this choice of
$\epsilon_{1\rm F}$ and $\nu_1$ ensures that, up to a displacement of the
energy origin, the oscillator potential arises for $q_1 = 1$, and also
that $\tilde V_1(x)$ is symmetric with respect to $x=0$. For $q_1\in(0,1)$
the potentials $\tilde V_1(x)$ have a double well, and the spacing between
the excited levels $E_n=n+1/2$ at $q_1=1$ is expanded by the factor
$q_1^{-2}$ while the ground state level is fixed at $\epsilon_{1\rm
F}=-1/2$.  On the other hand, for $q_1\in(1,\infty)$, the potentials
$\tilde V_1(x)$ present a single well centered at $x=0$, which can be
interpreted as a deformation of the oscillator bottom in order to maintain
fixed the ground state energy level at $\epsilon_{1\rm F}=-1/2$. The
excited levels $E_n=n+1/2$ at $q_1=1$ are `squeezed' by the factor
$q_1^{-2}$. 

It is important to remark once again the opposite physical mechanisms of
the spectral engineering taking place for the scaling intertwining (the
ground state is fixed and the excited levels are changing) and for the
standard intertwining technique (the ground state is changing and the
excited levels are fixed).

\hspace*{2truecm}
\begin{center}
\epsfxsize=9truecm
\epsfbox{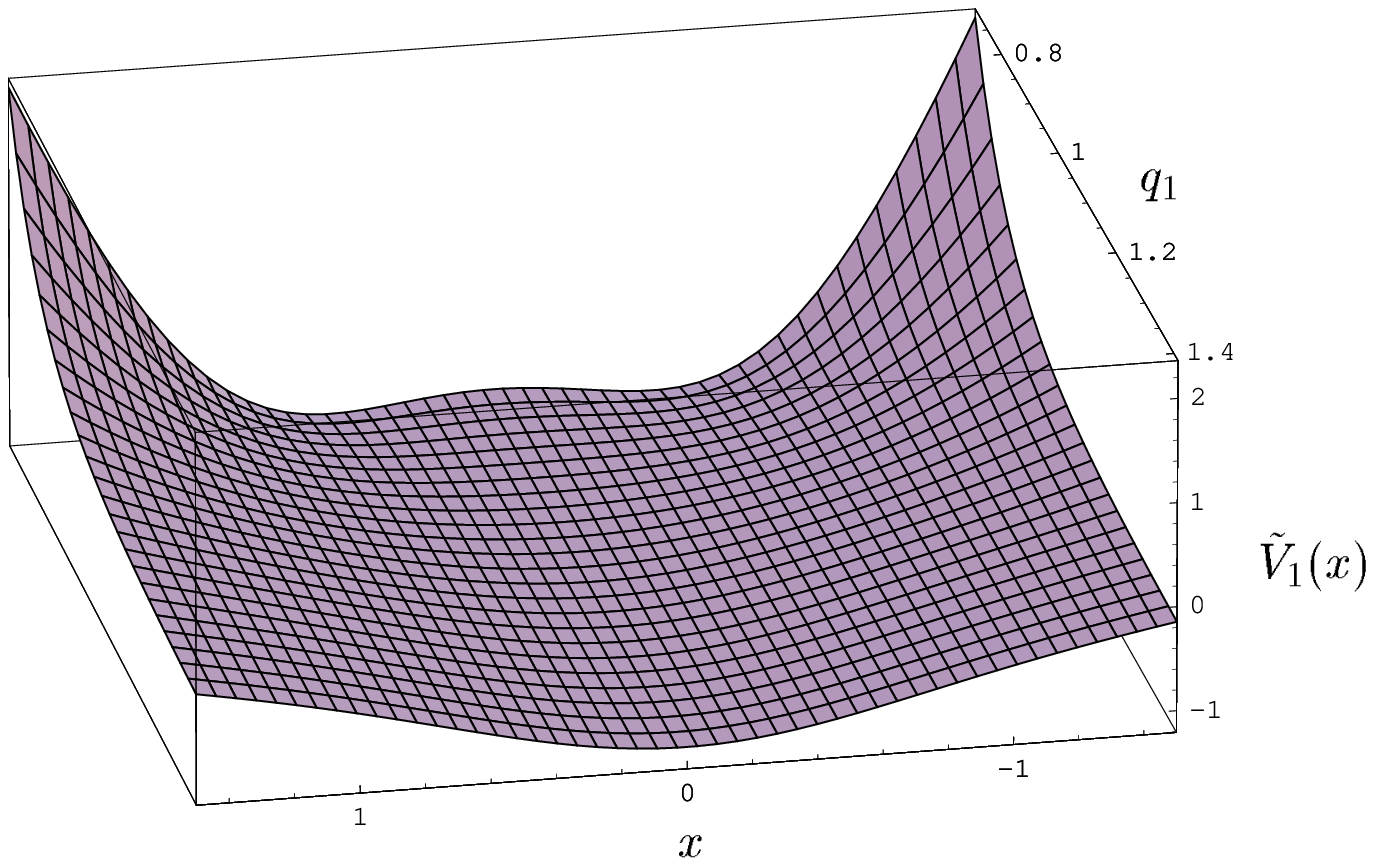}
\end{center}
\medskip
\begin{center}
{\small FIG. 3.
Spectral engineering of the potentials $\tilde V_1(x)$ of (38) as $q_1$
moves inside $[1/\sqrt{2},\sqrt{2}]$ with $\nu_1=0$ and $\epsilon_1 =
-q_1^2/2$. The ground state energy is fixed at $\epsilon _{1{\rm
F}}=-\frac12$, and the unscaled excited levels $E_n= n+ \frac12,
n=0,1,\dots$ are `expanded' or `squeezed' by the factor $q_1^{-2}$ for
$0<q_1<1$ or $q_1>1$, respectively.
}
\end{center} 

\medskip

\noindent {\bf (iv).} Let us discuss briefly the spectral manipulations
achievable by the iterative scaling intertwining of section 5.  The
potentials $\tilde V_2(x)$ after two iterations of the scaling
intertwining with $V_0(x) = x^2/2$ take the form
$$
\tilde V_2(x) = \frac1{(q_1q_2)^{2}}V_2 \left(\frac{x}{q_1q_2}\right) =   
(q_1q_2)^{-4}\frac{x^2}{2} 
$$
\begin{equation}
+\frac{1}{q_1q_2}\left[
\frac{2(\epsilon_1-\epsilon_2)}{\tilde{\alpha}_1(\frac{x}{q_1q_2},\epsilon_1)
- \tilde{\alpha}_1(\frac{x}{q_1q_2},\epsilon_2)}
\right]',
\end{equation}
where $\tilde{\alpha}_1(x,\epsilon_1) \equiv \alpha(x,\epsilon_1)$,
$\tilde{\alpha}_1(x,\epsilon_2) \equiv \alpha(x,\epsilon_2)$, $\epsilon_2
< \epsilon_1 < 1/2$, $\vert\nu_1\vert<1$ and $\vert\nu_2\vert>1$. The
spectrum of those potentials is of the form $\{(q_1q_2)^{-2}\epsilon_2,
(q_1q_2)^{-2}\epsilon_1, (q_1q_2)^{-2}(n+1/2), n=0,1,\dots\}$. Let us
notice that the potentials (39) form a six-parameter family labeled by
$\epsilon_1, \ q_1, \ \nu_1, \ \epsilon_2, \ q_2, \ \nu_2$, although the
spectra does dot depend on the values of $\nu_1$ and $\nu_2$ except for the
fact that they have to be taken in the domain for which $\tilde V_2(x)$
has no singularities. The presence of so many parameters provides us with
various possibilities for manipulating quantum spectra, as discussed at the 
end of section 5. For instance, if we take $\epsilon_1=q_1^2 \epsilon_{1\rm F}$,
with $\epsilon_{1\rm F}$, $q_2$ and $\epsilon_2$ fixed, by varying $q_1$
we will be manufacturing potentials with spectra of the type
$\{q_2^{-2}\epsilon_2, q_2^{-2}\epsilon_{1\rm F}, q_2^{-2}(n+1/2),
n=0,1,\dots\}$ scaled by the factor $q_1^{-2}$ excepting the first excited
level $q_2^{-2}\epsilon_{1\rm F}$, which is maintained fixed (see the
illustration of this effect on figure 4). Notice that this behaviour is
opposite to the one illustrated in figure 2 where the first excited level
is changing and the remaining part of the spectrum is fixed. 

\begin{center}
\hspace*{2truecm}
\epsfxsize=9truecm
\hskip-2.3cm\epsfbox{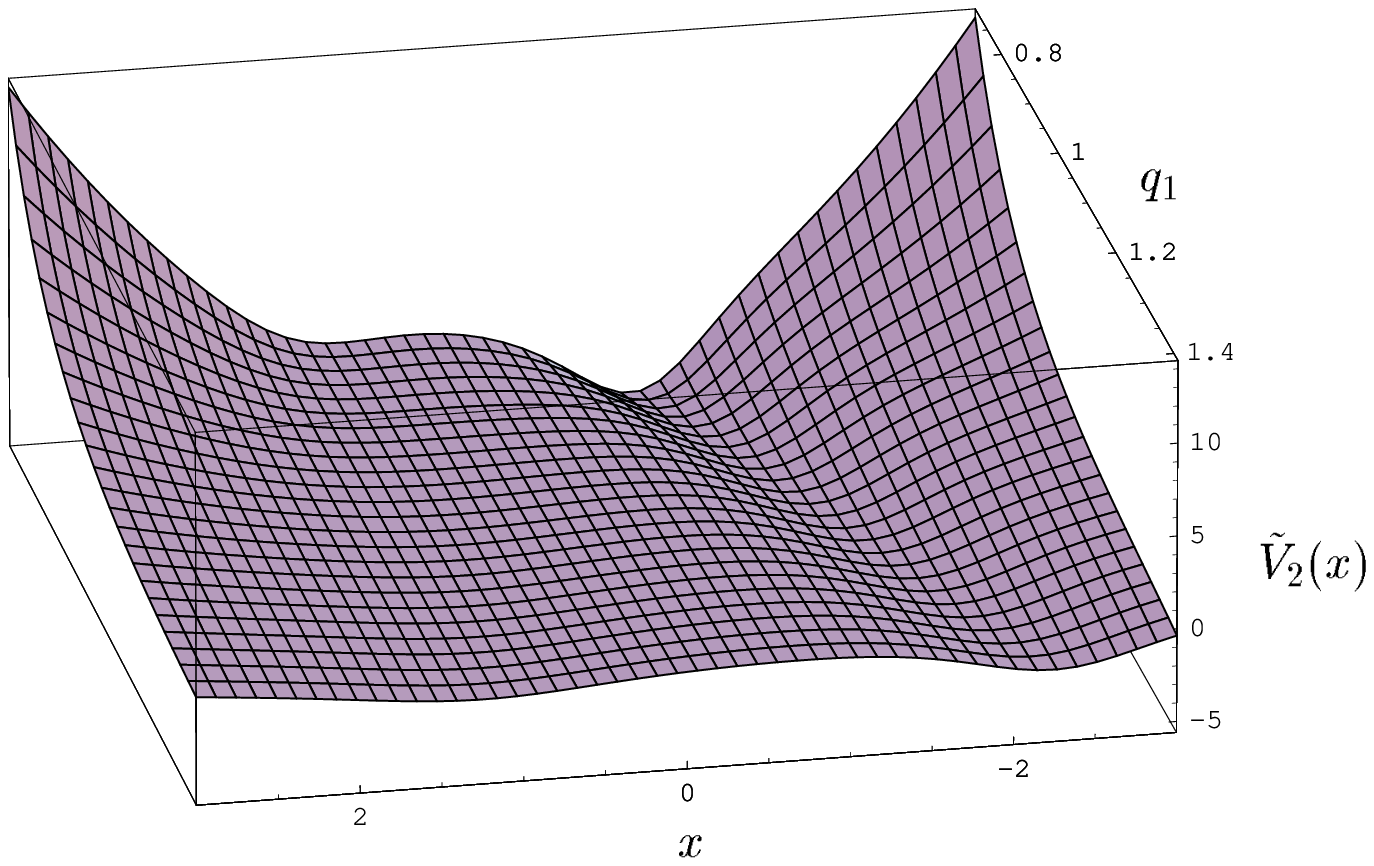}
\end{center}
\medskip
\begin{center}
{\small FIG. 4.
Spectral manipulation of $\tilde V_2(x)$ as $q_1$ moves inside
$[1/\sqrt{2},\sqrt{2}]$, $\nu_1=0, \ \nu_2 = 1.1$, $\epsilon_1 =
-\frac{q_1^2}{2}$, $\epsilon_2 = -\frac32$ and $q_2=1$. The first excited
level is fixed at $\epsilon_{1\rm F} = -\frac12$, while the other levels
$\{-\frac32,\frac12,\frac52,\dots\}$ are scaled by the factor
$q_1^{-2}$. 
}
\end{center}

A different possibility arises by taking $\epsilon_1 =
(q_1q_2)^{2}\epsilon_{1\rm F}, \ \epsilon_2 = (q_1q_2)^{2}\epsilon_{2\rm
F}$; when varying the product $q_1q_2$ we will be scaling the spectrum at
$q_1=q_2 =1$, $\{\epsilon_{1\rm F}, \epsilon_{2\rm F}, E_n = n+1/2\}$, by
the factor $(q_1q_2)^{-2}$ excepting the lowest two levels $\epsilon_{1\rm
F}$ and $\epsilon_{2\rm F}$ that remain static. An illustration of
this effect is presented in figure 5 for $\epsilon_{1\rm F}=\frac14$,
$\epsilon_{2\rm F} = -\frac12, \ \nu_1=0, \ \nu_2 = 10000, \ q_2=1$ and
$q_1 \in [\frac1{\sqrt{2}}, \sqrt{2})$.

\begin{center}
\hspace*{2truecm}
\epsfxsize=9truecm
\hskip-2.3cm\epsfbox{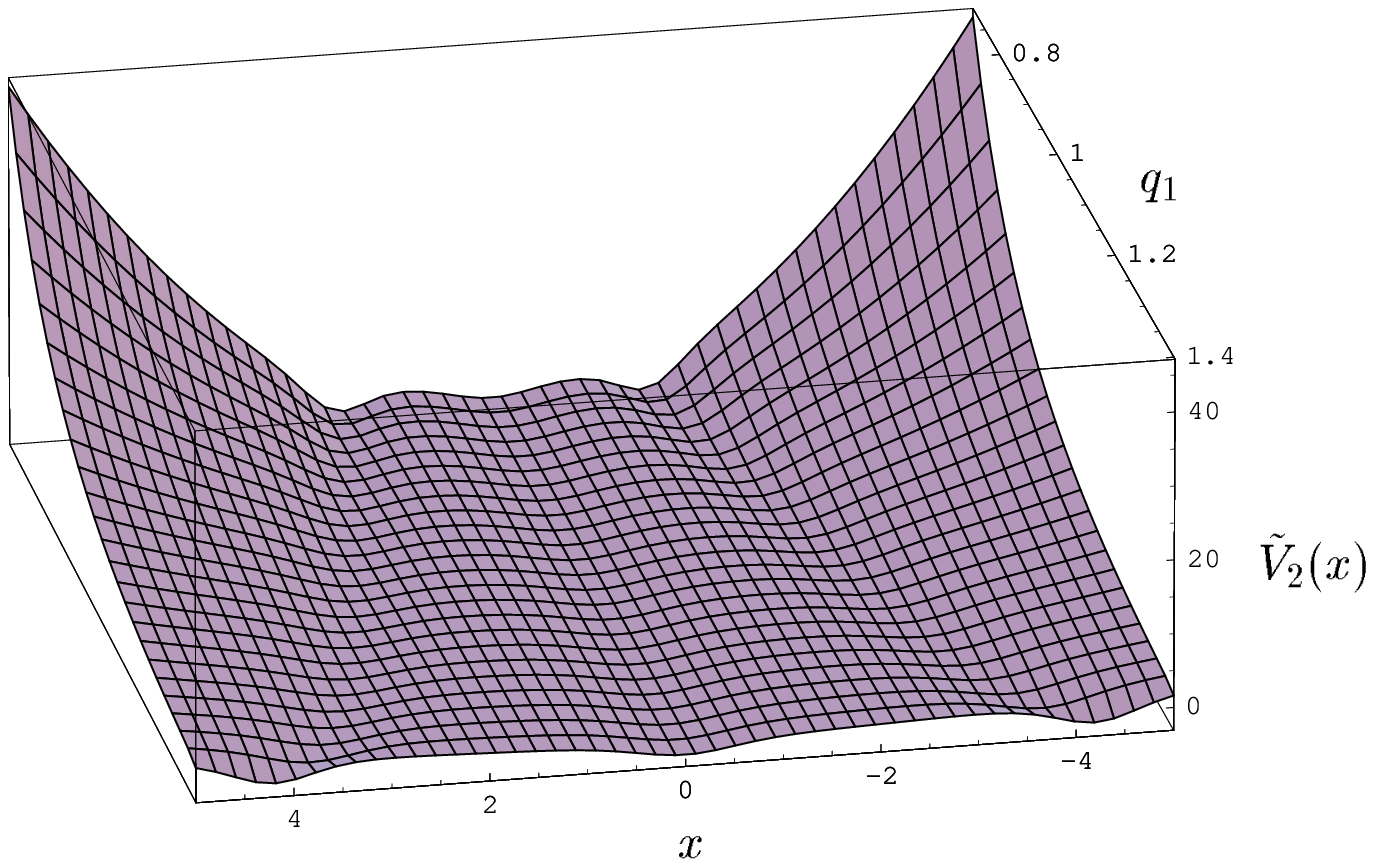}
\end{center}
\medskip
\begin{center}
{\small FIG. 5.
Spectral engineering of $\tilde V_2(x)$ for $q_1$ in
$[1/\sqrt{2},\sqrt{2})$, $\nu_1=0, \ \nu_2 = 10000$, $\epsilon_1 =
\frac{q_1^2}{4}$, $\epsilon_2 = -\frac{q_1^2}2$ and $q_2=1$. The two
lowest levels are fixed at $\epsilon_{1\rm F} = \frac14$ and
$\epsilon_{2\rm F} = -\frac12$, while the other ones $\{
\frac12,\frac32,\dots\}$ are scaled by the factor $q_1^{-2}$.
}
\end{center}

\section{Concluding remarks}
We presented some of the spectral engineering possibilities offered by the 
scaling intertwining procedure. Since this method is based on more parameters 
with respect to the standard ones
it can describe a wealth of technological and experimental situations. 
For this, it is sufficient 
to have a clear physical meaning of the parameters. For example, 
the scaling parameter can be associated to self-similar (fractal) properties
of micro and nanocavities, whereas the $\nu$ parameters are a measure of 
finite-size and boundary effects on the quantum spectra \cite{bef}.

Before closing, we mention that in interesting works, Spiridonov
\cite{spi92} has used a rather similar type of scaling intertwining
for emphasizing the connection between the $q$-deformed calculus and
Shabat's infinite chain of reflectionless potentials \cite{sha92}. In
such a treatment the so-called self-similar potentials have been
introduced and characterized, and at the same time the corresponding
spectrum has been determined by means of the algebraic technique itself. 
On the other hand, in this paper the orientation has been different, namely to
take initially a potential with known spectrum in order to generate a new
potential with known and different (in general) spectrum. When we make
use of the freedom provided by the different parameters of the families of
potentials generated by means of the various intertwinings (standard
and/or scaled ones), we arrive at the spectral engineering process widely
discussed in this paper.

\section{Acknowledgements}

\noindent This work was supported in part by the CONACyT Projects 32086-E
and 458100-5-25844E. HCR wishes to thank for the kind hospitality at the
Departamento de F\'{\i}sica, CINVESTAV.



\end{document}